\documentclass[twocolumn,prb,showpacs,floatfix]{revtex4}

\usepackage{graphicx}
\usepackage{dcolumn}
\usepackage{bm}
\usepackage{amsmath}

\begin{document}

\title{
From a few to many electrons in quantum dots under strong magnetic fields:\\
Properties of rotating electron molecules with multiple rings
}

\author{Yuesong Li}
\author{Constantine Yannouleas}
\author{Uzi Landman}

\affiliation{School of Physics, Georgia Institute of Technology,
             Atlanta, Georgia 30332-0430}

\date{08 December 2005}

\begin{abstract}
Using the method of breaking of circular symmetry and of subsequent symmetry 
restoration via projection techiques, we present calculations for the 
ground-state energies and excitation spectra of $N$-electron parabolic quantum 
dots in strong magnetic fields in the medium-size range $10 \leq N \leq 30$. The 
physical picture suggested by our calculations is that of finite 
rotating electron molecules (REMs) comprising multiple rings, 
with the rings rotating independently of each other. An analytic expression 
for the energetics of such non-rigid multi-ring REMs is derived; 
it is applicable to arbitrary sizes given the corresponding equilibrium 
configuration of classical point charges. We show that the rotating electron 
molecules have a non-rigid (non-classical) rotational inertia exhibiting 
simultaneous crystalline correlations and liquid-like (non-rigidity)
characteristics. This mixed phase appears in high magnetic fields and contrasts 
with the picture of a classical rigid Wigner crystal in the lowest Landau level.
\end{abstract}

\pacs{73.21.La; 71.45.Gm; 71.45.Lr}

\maketitle

\section{Introduction}
\subsection{Computational motivation}

Due to the growing interest in solid-state nanostructures, driven by basic
research and potential technological considerations, two-dimensional 
$N$-electron semiconductor quantum dots (QDs) in field-free conditions and under
applied magnetic fields ($B$) 
have been extensively studied in the last few years, both experimentally 
\cite{kou,cior,marc} and theoretically. \cite{maks,ruan2,ruan,yl,pee,yl2}
Experimentally, the case of parabolic QDs with a small number of electrons 
($N \leq 30$) has attracted particular attention, as a result of precise 
control of the number of electrons in the dot that has been demonstrated
in several experimental investigations.

Naturally, QDs with a small number of electrons are 
also most attractive for theoretical 
investigations, since their ground-state properties and excitation 
spectra can be analyzed \cite{maks,ruan2,ruan,yl,pee,yang,haw2,haw1} through 
exact-diagonalization (EXD) solutions of the many-body Schr\"{o}dinger equation. 
In particular, in combination with certain approximate methods, 
which are less demanding computationally 
while providing highly accurate results and a transparent 
physical picture (e.g., the method of successive hierarchical approximations, 
\cite{yl,yl2} see below), EXD calculations confirmed the 
spontaneous formation of finite rotating electron molecules (REMs) 
and the description of the excited states with magic angular momenta as 
yrast rotational bands of these REMs \cite{yl} (sometime the REMs are
referred to as ``rotating Wigner molecules,'' RWMs). However, the number of 
Slater determinants in the EXD wave-function expansion
increases exponentially as a function of $N$, and as a result
EXD calculations to date have been restricted to rather low values of $N$, 
typically with $N \lesssim 10$; this has prohibited investigation of REMs
with multiple rings. A similar problem appears also with other 
wave functions that are expressed as a discrete sum over Slater 
determinants, such as the {\it analytic\/} REM wave functions
[7(a,b)], or the variational Monte Carlo approach of Ref.\ \onlinecite{sil}.

Most EXD calculations (see, e.g., Refs.\ \onlinecite{maks}, 
\onlinecite{yl}(b), \onlinecite{yang,haw2,nie}) have been carried out in the 
regime of very strong magnetic field (i.e., $B \rightarrow \infty$), 
such that the Hilbert space can be restricted to the lowest Landau level (LLL);
in this regime, the confinement does not have any influence on the 
composition of the microscopic many-body wave function (see section II.B).   
EXD calculations as a function of $B$ that include explicitly the full effect of
the confinement, \cite{ruan2,ruan,pee,haw1} i.e., mixing with higher Landau 
levels are more involved, and thus they are scarce and are usually restricted to
very small sizes with $N \leq 4$. An exception is presented
by the method of hyperspherical harmonics, \cite{ruan2,ruan} which, however,
may not be reliable for all the sizes up to $N \sim 10$ (see below).

Systematic EXD calculations beyond the numerical barrier of $N \sim 10 $ 
electrons are not expected to become feasible in the near future. In this paper,
we show that a microscopic numerical method, which was introduced by us
recently and is based on successive hierarchical approximations (with 
increasing accuracy after each step) is able to go beyond this barrier. 
This approach (referred to, for brevity, as the ``two-step method'') can provide 
high-quality calculations describing properties of QDs as a function of $B$ in 
the whole size range $2 \leq N \leq 30$, {\it with (or without) consideration 
of the effect of the confinement on the mixing with higher Landau levels.\/} 
In this paper, we will consider the case of fully polarized electrons, which 
in typical GaAs experimental devices is appropriate for strong $B$ such 
that the ground-state angular momentum $L \geq L_0 \equiv N(N-1)/2$ (see section
II.A and footnotes therein). 

The minimum value $L_0$ specifies the so-called 
maximum density droplet (MDD); its name results from the fact that it was
originally defined \cite{mac31} in the LLL where it is a single Slater 
determinant built out of orbitals with contiguous single-particle angular 
momenta 0,1,2, ..., $N-1$. We will show, however, (see Section IV.B) that 
mixing with higher Landau levels is non-negligible for MDD ground
states that are feasible in currently available experimental quantum dots;
in this case the electron density of the MDD is not constant, but 
it exhibits oscillations. 

\subsection{Nonclassical (non-rigid) rotational inertia}

The existence of an exotic supersolid crystalline phase with combined solid and
superfluid characteristics has been long conjectured \cite{anlf,ches,legg}
for solid $^4$He under appropriate conditions. The recent experimental
discovery \cite{chan} that solid $^4$He exhibits a nonclassical (nonrigid)
rotational inertia (NCRI \cite{legg}) has revived an intense interest
\cite{legg2,datt,prok,cepe,fala} in the existence and properties of the
supersolid phase in this system, as well as in the possible emergence of
exotic phases in other systems.

As we show here, certain aspects of supersolid behavior (e.g., the
simultaneous occurrence of crystalline correlations and non-rigidity under
rotations) may be found for electrons in quantum dots.
As aforementioned, under a high magnetic field, the electrons confined in a QD 
localize at the vertices of concentric polygonal rings and form a rotating 
electron molecule. \cite{yl} We show that the
corresponding rotational inertia strongly deviates from the rigid classical
value, a fact that endows the REM with supersolid-like characteristics
(in the sense of the appearance of a non-classical rotational inertia, but
without implying the presence of a superfluid component). 
Furthermore, the REM at high $B$ can be naturally viewed as the precursor of a 
quantum crystal that develops in the lowest Landau level (LLL) in the 
thermodynamic limit. Due to the lack of rigidity, the LLL quantum crystal 
exhibits  a ``liquid''-like behavior. These conclusions were
enabled by the development of an analytic expression for the excitation 
energies of the REM that permits calculations for an arbitrary number of 
electrons, given the classical polygonal-ring structure in the QD. \cite{kon}

The paper is organized as follows. Section II is devoted to a description
of computational methods for the properties of electrons in QDs under
high magnetic fields, with explicit consideration of effects due to
the external confinement. In section III, we compare results from various
computationals methods with those obtained via exact diagonalization.
Illustrative examples of the formation of crystalline rotating electron
molecules with ground-state multiple concentric polygonal ring structures,
and their isomers, are given in section IV for QDs with $N=6$, 9, 11, 17.
The yrast band of rotational excitations (at a given $B$) is analyzed 
in section V along with the derivation of an analytic formula that provides
for stronger fields (and/or higher angular momenta) accurate predictions of the 
energies of REMs with arbitrary numbers of electrons. In section VI, we discuss
the non-rigid (liquid-like) characteristics of electrons in quantum dots
under high magnetic fields as portrayed by their non-classical rotational 
inertia. We summarize our findings in section VII.
For an earlier shorter version of this paper, see Ref.\ \onlinecite{cond}.

\section{Description of computational methods that consider the external
confinement}

\subsection{The REM microscopic method}

In our method of successive hierarchical approximations, we begin with a 
{\it static\/} electron molecule (SEM), described by an unrestricted 
Hartree-Fock (UHF) determinant that violates the circular symmetry. \cite{yl2}
Subsequently, the {\it rotation\/} of the electron molecule is described by a
post-Hartree-Fock step of restoration of the broken circular symmetry via
projection techniques. \cite{yl,yl2} 
Since we focus here on the case of strong $B$, we can approximate
the UHF orbitals (first step of our procedure) by (parameter free) displaced
Gaussian functions; that is, for an electron localized at ${\bf R}_j$
($Z_j$), we use the orbital
\begin{equation}
u(z,Z_j) = \frac{1}{\sqrt{\pi} \lambda}
\exp \left( -\frac{|z-Z_j|^2}{2\lambda^2} - i\varphi(z,Z_j;B) \right),
\label{uhfo}
\end{equation}
with $\lambda = \sqrt{\hbar /m^* \Omega}$;
$\Omega=\sqrt{\omega_0^2+\omega_c^2/4}$, where $\omega_c=eB/(m^*c)$ is the
cyclotron frequency and $\omega_0$ specifies the external parabolic confinement.
We have used complex numbers to represent the position
variables, so that $z=x+iy$, $Z_j = X_j +i Y_j$.
The phase guarantees gauge invariance in the presence of
a perpendicular magnetic field and is given in the symmetric gauge by
$\varphi(z,Z_j;B) = (x Y_j - y X_j)/2 l_B^2$, with $l_B = \sqrt{\hbar c/ e B}$.

For an extended 2D system, the $Z_j$'s form a triangular lattice.
\cite{yosh} For finite $N$, however, the $Z_j$'s coincide \cite{yl,yl2,yl3}
with the equilibrium positions [forming $r$ concentric
regular polygons denoted as ($n_1, n_2,...,n_r$)] of $N=\sum_{q=1}^r n_q$
classical point charges inside an external parabolic confinement. \cite{kon}
In this notation, $n_1$ corresponds to the innermost ring with $n_1 > 0$. 
For the case of a single polygonal ring, the notation $(0,N)$ is often used;
then it is to be understood that $n_1=N$.

The wave function of the {\it static\/} electron molecule (SEM) is a
{\it single\/} Slater determinant $|\Psi^{\text{SEM}} [z] \rangle$ made out of
the single-electron wave functions $u(z_i,Z_i)$, $i = 1,...,N$.
Correlated many-body states with good total angular momenta $L$
can be extracted \cite{yl} (second step) from the UHF determinant using 
projection operators. The projected rotating electron molecule state
is given by
\begin{eqnarray}
|\Phi^{\text{REM}}_L \rangle & = & \int_0^{2\pi} ... \int_0^{2\pi}
d\gamma_1 ... d\gamma_r \nonumber \\
&& \times |\Psi^{\text{SEM}}(\gamma_1, ..., \gamma_r) \rangle
\exp \left( i \sum_{q=1}^r \gamma_q L_q \right).
\label{wfprj}
\end{eqnarray}
Here $L=\sum_{q=1}^r L_q$ and
$|\Psi^{\text{SEM}}[\gamma] \rangle$ is the original Slater determinant
with {\it all the single-electron wave functions of the $q$th ring\/} rotated
(collectively, i.e., coherently) by the {\it same\/} azimuthal angle $\gamma_q$.
Note that Eq.\ (\ref{wfprj}) can be written as a product of projection
operators acting on the original Slater determinant [i.e., on
$|\Psi^{\text{SEM}}(\gamma_1=0, ..., \gamma_r=0) \rangle$].
Setting $\lambda = l_B \sqrt{2}$ restricts the single-electron wave function in
Eq. (\ref{uhfo}) to be entirely in the lowest Landau level \cite{yl}
(see Appendix A). 
The continuous-configuration-interaction form of the projected wave functions
[i.e., the linear superposition of determimants in Eq.\ (\ref{wfprj})]
implies a highly entangled state. We require here that $B$ is sufficiently
strong so that all the electrons are spin-polarized \cite{note42} 
and that the ground-state angular momentum $L \geq L_0 \equiv N(N-1)/2$ 
(or equivalently that the fractional filling factor $\nu \equiv L/L_0 \leq 1$).

Due to the point-group symmetries of each polygonal ring of electrons
in the SEM wave function, the total angular momenta $L$ of the rotating
crystalline electron molecule are restricted to the so-called {\it magic} 
angular momenta, i.e.,
\begin{equation}
L_m = L_0 + \sum_{q=1}^r k_q n_q,
\label{lmeq}
\end{equation}
where the $k_q$'s are non-negative integers\cite{note44}
(when $n_1=1$, $k_1=0$).

The partial angular momenta associated with the $q$th ring,
$L_q$ [see Eq.\ (\ref{wfprj})], are given by
\begin{equation}
L_q = L_{0,q} + k_q n_q,
\label{lmpar}
\end{equation}
where $L_{0,q}=\sum_{i=i_q+1}^{i_q+n_q} (i-1)$ with
$i_q = \sum_{s=1}^{q-1} n_s$ $(i_1=0)$, and
$L_0 = \sum_{q=1}^r L_{0,q}$.

The energy of the REM state [Eq.\ (\ref{wfprj})] is given\cite{yl2,yl3} by
\begin{equation}
E^{\text{REM}}_L = \left. { \int_0^{2\pi} h([\gamma]) e^{i [\gamma] \cdot [L]}
d[\gamma] } \right/%
{ \int_0^{2\pi} n([\gamma]) e^{i [\gamma] \cdot [L]} d[\gamma]},
\label{eproj}
\end{equation}
with the hamiltonian and overlap matrix elements $h([\gamma]) =
\langle \Psi^{\text{SEM}}([0]) | H | \Psi^{\text{SEM}}([\gamma]) \rangle$ and
$n([\gamma]) =
\langle \Psi^{\text{SEM}}([0]) | \Psi^{\text{SEM}}([\gamma]) \rangle$, 
respectively, and
$[\gamma] \cdot [L] = \sum_{q=1}^r \gamma_q L_q$.
The SEM energies are simply given by $E_{\text{SEM}} = h([0])/n([0])$.

The many-body Hamiltonian is
\begin{equation}
H = \sum_{i=1}^N H_{\text{sp}}(i) + 
\sum_{i=1}^N \sum_{j > i}^N \frac{e^2}{\kappa r_{ij}},
\label{mbh}
\end{equation}
with 
\begin{equation}
H_{\text{sp}}(i)=
\frac{1}{2m^*}\left( {\bf p}_i- \frac{e}{c} {\bf A}_i \right)^2
+ \frac{m^*}{2} \omega_0^2 {\bf r}_i^2,
\label{hsp}
\end{equation}
being the single-particle part. The hamiltonian $H$ 
describes $N$ electrons (interacting via a Coulomb repulsion) confined by
a parabolic potential of frequency $\omega_0$ and subjected to
a perpendicular magnetic
field $B$, whose vector potential is given in the symmetric gauge by
${\bf A(r)} = \frac{1}{2} (-By,Bx,0)$.
$m^*$ is the effective electron mass, $\kappa$ is the  dielectric
constant of the semiconductor material, and $r_{ij}=|{\bf r}_i - {\bf r}_j|$.
For sufficiently high magnetic fields, the electrons are fully spin-polarized 
and the Zeeman term [not shown in Eq.\ (\ref{mbh})] does not need to be
considered. \cite{note42} Thus the calculations in this paper do not include
the Zeeman contribution, which, however, can easily be added (for a fully
polarized dot, the Zeeman contribution to the total energy is 
$Ng^* \mu_B B/2$, with $g^*$ being the effective Land\'{e} factor and 
$\mu_B$ the Bohr magneton).

The crystalline polygonal-ring arrangement $(n_1,n_2,...,n_r)$ of classical
point charges is portrayed directly in
the electron density of the broken-symmetry SEM, since the latter consists of
humps centered at the localization sites $Z_j$'s ({\it one hump} for each
electron). In contrast, the REM has good angular momentum and thus its electron
density is circularly uniform. To probe the crystalline character of the REM,
we use the conditional probability distribution (CPD) defined as
\begin{equation}
P({\bf r},{\bf r}_0) =
\langle \Phi |
\sum_{i \neq j}  \delta({\bf r}_i -{\bf r})
\delta({\bf r}_j-{\bf r}_0)
| \Phi \rangle  / \langle \Phi | \Phi \rangle,
\label{cpds}
\end{equation}
where $\Phi ({\bf r}_1, {\bf r}_2, ..., {\bf r}_N)$
denotes the many-body wave function under consideration.
$P({\bf r},{\bf r}_0)$ is proportional to the conditional probability of
finding an electron at ${\bf r}$, given that another electron is assumed at 
${\bf r}_0$. This procedure subtracts the collective rotation of the electron 
molecule in the laboratory frame of referenece, and, as a result, the
CPDs reveal the structure of the many body state in the intrinsic (rotating)
reference frame.

\subsection{Exact diagonalization in the lowest Landau level} 

We describe here a widely used approximation\cite{maks,yang,jeon}
for calculating the ground state at a given $B$, which takes advantage of 
the simplifications at the $B \rightarrow \infty$ limit, i.e., when the
relevant Hilbert space can be restricted to the lowest Landau level 
[then $\hbar \omega_0 << \hbar \omega_c /2$ (for $B \rightarrow \infty$) and 
the confinement can be neglected at a first step].
Then, the many-body hamiltonian [see Eq.\ (\ref{mbh})] reduces to
\begin{equation}
H^{B \rightarrow \infty}_{\text{LLL}} = N \frac{\hbar \omega_c}{2} 
+ \sum_{i=1}^N \sum_{j > i}^N \frac{e^2}{\kappa r_{ij}}.
\label{hlll}
\end{equation}

Due to the form of the limiting Hamiltonian in Eq.\ (\ref{hlll}), 
one can overlook the zero-point-energy term and perform an
exact diagonalization only for the Coulomb interaction part. 
The corresponding interaction energies can be written as 
\begin{equation}
\widetilde{E}^{\text{EXD}}_{\text{int,LLL}} (L) = 
\widetilde{{\cal E}}^{\text{EXD}}_{\text{int,LLL}} (L) \frac{e^2}{\kappa l_B},
\label{exlll}
\end{equation}
where $\widetilde{{\cal E}}^{\text{EXD}}_{\text{int,LLL}}$ is dimensionless.
The subscript ``int'' identifies the $e-e$ interaction as the source of this
term.

In this approximation scheme,
at finite $B$ the external confinement $\hbar \omega_0$ is taken
into consideration only through the lifting of the single-particle
degeneracy within the LLL, while disregarding higher Landau levels.
As a result, the effect of the confinement enters here only as follows: 
(I) in the interaction term [see Eq.\ (\ref{exlll})], one scales the effective 
magnetic length, i.e., one replaces $l_B$ by $\lambda / \sqrt{2}$ 
(see section II.A for the definition of $l_B$ and $\lambda$) without modifying 
the dimensionless part 
$\widetilde{{\cal E}}^{\text{EXD}}_{\text{int,LLL}}$, and (II) the 
contribution, $E^{n=0}_{\text{sp}}(B,L)$ (referenced to $N \hbar \Omega$), 
of the single-particle hamiltonian $\sum_{i=1}^N H_{\text{sp}}(i)$ 
to the total energy [see Eq.\ (\ref{mbh})] is added to 
$\widetilde{E}^{\text{EXD}}_{\text{int,LLL}} (L)$ 
[corresponding to the second term on the right-hand side of Eq. (\ref{mbh})]. 
$E^{n=0}_{\text{sp}}(B,L)$ is the sum 
of Darwin-Fock single-particle energies $\epsilon^{\text{DF}}_{n,l}$ with zero 
nodes ($n=0$; the corresponding single-particle 
states become degenerate at $B \rightarrow \infty$ and form the lowest Landau 
level). Since 
\begin{equation}
\epsilon^{\text{DF}}_{n,l} = (2n+1+|l|)\hbar \Omega - l \hbar \omega_c /2,
\label{enldf}
\end{equation}
the $E^{n=0}_{\text{sp}}(B,L)$ is linear in the total angular momentum 
$L=\sum_{i=1}^N l_i$, i.e.,
\begin{equation}
E^{n=0}_{\text{sp}}(B,L) = \hbar (\Omega -\omega_c/2)L.
\label{n0esp}
\end{equation}
Note that $E^{n=0}_{\text{sp}}(B \rightarrow \infty ,L) \rightarrow 0$.

We denote the final expression of this approximation by 
$\widetilde{E}^{\text{EXD}}_{\text{tot,LLL}}$; it is given by
\begin{equation}
\widetilde{E}^{\text{EXD}}_{\text{tot,LLL}} (B,L) = 
E^{n=0}_{\text{sp}}(B,L)
+ \sqrt{2} \widetilde{{\cal E}}^{\text{EXD}}_{\text{int,LLL}} (L) 
\frac{e^2}{\kappa \lambda}.
\label{limexd}
\end{equation}

An approximate ground-state energy for the system can be found through Eq.\
(\ref{limexd}) by determining the angular-momentum value $L_{\text{gs}}$ that 
minimizes this expression. In the following, this ground-state energy at a given
$B$ will be denoted simply by omitting the variable $L$ on the left-hand-side
of Eq.\ (\ref{limexd}), i.e.,
$\widetilde{E}^{\text{EXD}}_{\text{tot,LLL}} (B) \equiv 
\widetilde{E}^{\text{EXD}}_{\text{tot,LLL}} (B,L_{\text{gs}})$.   

We note that, although few in number, full EXD calculations for finite $B$ 
that take into consideration both the confinement $\hbar \omega_0$ and the 
actual complexity of the Darwin-Fock spectra (including levels with $n > 0$) 
have been reported \cite{ruan2,ruan,pee,haw1} in the literature for several 
cases with $N=3$ and $N=4$ electrons. These calculations will be of great 
assistance in evaluating the accuracy of the REM method (see section III). 

In the above Eq. (\ref{limexd}), we have used exact diagonalization in the
lowest Landau level for evaluating the interelectron interaction 
contribution to the total energy. In alternative treatments, one may obtain
the interelectron energy contribution through the use of various approximate
wave functions restricted to the LLL. These include the use of the Laughlin
wave function and descendants thereof (e.g., composite fermions), or the
rotating electron wave functions at the limit $B \rightarrow \infty$, which
is reached by setting $\lambda = l_B \sqrt{2}$ in the right-hand-side of
Eq. (\ref{uhfo}) (defining the displaced orbital).
For these cases, we will use the obvious notations
$\widetilde{E}^{\text{Laughlin}}_{\text{tot,LLL}} (B,L)$,
$\widetilde{E}^{\text{CF}}_{\text{tot,LLL}} (B,L)$, and
$\widetilde{E}^{\text{REM}}_{\text{tot,LLL}} (B,L)$.

\begin{figure}[t]
\centering\includegraphics[width=6.3cm]{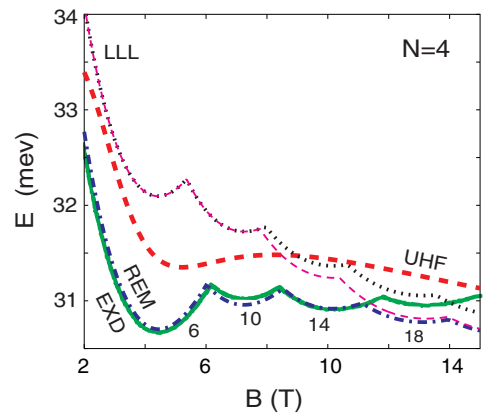}
\caption{(Color online)
Two-step-method versus EXD calculations: Ground-state energies for $N=4$
electrons (referenced to $4 \hbar \Omega$) as a function of the 
magnetic field $B$.
Thick dashed line (red): broken-symmetry UHF (SEM).
Solid line (green): EXD (from Ref.\ \onlinecite{ruan2}).
Thick dashed-dotted line (blue): REM.
Thin dashed line (violet, marked LLL): the commonly used approximate energies
$\widetilde{E}^{\text{EXD}}_{\text{tot,LLL}}(B)$ [see Eq.\ (\ref{limexd})]. 
Thin dotted line (black): $\widetilde{E}^{\text{REM}}_{\text{tot,LLL}}(B)$ 
(see section II.B).
For $ B < 8$ T, the $\widetilde{E}^{\text{EXD}}_{\text{tot,LLL}}(B)$ and
$\widetilde{E}^{\text{REM}}_{\text{tot,LLL}}(B)$ curves coincide;
we have checked that these curves approach each other also at larger
values of $B$, outside the plotted range. Numbers near the bottom curves
denote the value of magic angular momenta [$L_m$, see Eq.\ (\ref{lmeq})]
of the ground state. Corresponding fractional filling factors are specified by
$\nu = N(N-1)/(2L_m)$.
Parameters used: confinement $\hbar \omega_0=3.60$ meV, 
dielectric constant $\kappa=13.1$, effective mass $m^*=0.067 m_e$. 
}
\end{figure}

\section{Comparison of approximate results with exact diagonalization 
calculations}

\subsection{Ground-state energies in external confinement}

Before proceeding with the presentation of results for $N > 10$,
we demonstrate the accuracy of the two-step method through comparisons with
existing EXD results for smaller sizes.  
In Fig.\ 1, our calculations for ground-state energies as a function of $B$ are 
compared to EXD calculations \cite{ruan2} for $N=4$ electrons in an external 
parabolic confinement. The thick dotted line (red) represents the 
broken-symmetry UHF approximation (first step of our method), 
which naturally is a smooth curve lying above the EXD one [solid line (green)]. 
The results obtained after restoration of symmetry [dashed-dotted line (blue); 
marked as REM] agree very well with the EXD one in the whole range 
2 T$ < B <$ 15 T. \cite{note1} We recall here that, for the parameters of the
QD, the electrons form in the intrinsic frame of reference a square about
the origin of the dot, i.e., a (0,4) configuration, with the zero indicating
that no electron is located at the center. According to Eq.\ (\ref{lmeq}),
$L_0=6$, and the magic angular momenta are given by $L_m=6+4k$,
$k=0$, 1, 2, ...

\begin{figure}[b]
\centering\includegraphics[width=6.8cm]{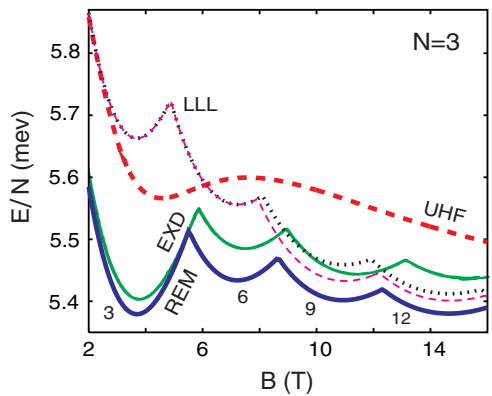}
\caption{(Color online)
Two-step method versus EXD calculations: Ground-state energies (per particle, 
referenced to $\hbar \Omega$) for $N=3$ electrons. 
The electrons are arranged in a (0,3) structure in the intrinsic frame of
reference. Thick dashed line (red): broken-symmetry UHF (SEM). 
Thinner solid line (green): EXD (from Ref.\ \onlinecite{haw1}).
Thick solid line (blue): REM.
Thin dashed line (violet): the commonly used approximate energies  
$\widetilde{E}^{\text{EXD}}_{\text{tot,LLL}}(B)$ (see text). Thin dotted line 
(black): $\widetilde{E}^{\text{REM}}_{\text{tot,LLL}}(B)$ (see text). 
For $ B < 8$ T, the $\widetilde{E}^{\text{EXD}}_{\text{tot,LLL}}(B)$ and
$\widetilde{E}^{\text{REM}}_{\text{tot,LLL}}(B)$ curves coincide;
we have checked that these curves approach each other also at larger
values of $B$, outside the plotted range.
Numbers denote the value of magic angular momenta ($L_m$) of the ground state.
Corresponding fractional filling factors are specified by
$\nu = N(N-1)/(2L_m)$.
Parameters used: confinement $\hbar \omega_0=3.37$ meV, 
dielectric constant $\kappa=12.4$, effective mass $m^*=0.067 m_e$. 
}
\end{figure}

To further evaluate the accuracy of the two-step method, we also display 
in Fig.\ 1 [thin dashed line
(violet)] ground-state energies calculated with the commonly used 
\cite{maks,yang,jeon} approximation 
$\widetilde{E}^{\text{EXD}}_{\text{tot,LLL}}(B)$ 
(see section II.B). We find that the energies 
$\widetilde{E}^{\text{EXD}}_{\text{tot,LLL}} (B)$ 
tend to substantially overestimate
the REM (and EXD) energies for lower values of $B$ (e.g., by as much as 5.5\% at 
$B \sim 4$ T). On the other hand, for higher values of $B$ ($>$ 12 T), the 
energies $\widetilde{E}^{\text{EXD}}_{\text{tot,LLL}}(B)$ 
tend to agree rather well with 
the REM ones.  A similar behavior is exhibited also by the 
$\widetilde{E}^{\text{REM}}_{\text{tot,LLL}}(B)$ 
energies [the interaction energies are 
calculated within the LLL using the REM wave function; dotted line (black)].
We have found that the overestimation exhibited by the 
$\widetilde{E}^{\text{EXD}}_{\text{tot,LLL}}(B)$ 
energies is due to the fact that the actual
dimensionless Coulomb coefficient 
$\widetilde{{\cal E}}^{\text{EXD}}_{\text{int,LLL}} (L)$
[See Eq.\ (\ref{limexd})] is not independent of the magnetic field, 
but decreases slowly as $B$ decreases when the effect of the 
confinement is considered (see Appendix B). A similar agreement 
between REM and EXD results, and a similar inaccurate behavior of the 
limiting-case approximation, was found by us also for $N=3$ electrons in the 
range 2 T $< B <$ 16 T shown in Fig.\ 2 (the EXD calculation was taken from 
Ref.\ \onlinecite{haw1}). 

In all cases, the total energy of the REM is lower than that of the SEM 
(see, e.g., Figs.\ 1 and 2). Indeed, a theorem discussed in Sec. III of 
Ref. \onlinecite{low}, pertaining
to the energies of projected wave functions, guarantees that this lowering of
energy applies for all values of $N$ and $B$.

\subsection{Yrast rotational band at $B \rightarrow \infty$}

As a second accuracy test, we compare in Table I REM and EXD results for the
interaction energies of the yrast band for $N=6$ electrons in the lowest Landau
level [an yrast state is the lowest energy state for a given magic angular
momentum $L_m$, Eq.\ (\ref{lmeq})]. The relative error is smaller than 0.3\%, 
and it decreases steadily for larger $L$ values. 

\begin{table}[t]
\caption{%
Comparison of yrast-band energies obtained from REM and EXD calculations for
$N=6$ electrons in the lowest Landau level, that is in the
limit $B \rightarrow \infty$. In this limit the external confinement can
be neglected and only the interaction energy contributes to the yrast-band
energies. Energies in units of $e^2/(\kappa l_B)$. For the REM results, the
(1,5) polygonal-ring arrangement was considered. For $L < 140$, see
Table IV in Ref.\ \onlinecite{yl}(b) and Table III in Ref.\ \onlinecite{yl2}(c).
The values of the fractional filling may be obtained for each $L$ as
$\nu=N(N-1)/(2L)$.
}
\begin{ruledtabular}
\begin{tabular}{rccc}
$L$  & REM & EXD  & Error (\%)   \\ \hline
140  & 1.6059  & 1.6006  & 0.33 \\
145  & 1.5773  & 1.5724  & 0.31 \\
150  & 1.5502  & 1.5455  & 0.30 \\
155  & 1.5244  & 1.5200  & 0.29 \\
160  & 1.4999  & 1.4957  & 0.28 \\
165  & 1.4765  & 1.4726  & 0.27 \\
170  & 1.4542  & 1.4505  & 0.26 \\
175  & 1.4329  & 1.4293  & 0.25 \\
180  & 1.4125  & 1.4091  & 0.24 \\
185  & 1.3929  & 1.3897  & 0.23 \\
190  & 1.3741  & 1.3710  & 0.23 \\
195  & 1.3561  & 1.3531  & 0.22 \\
200  & 1.3388  & 1.3359  & 0.21
\end{tabular}
\end{ruledtabular}
\end{table}

\section{Illustrative examples from microscopic REM calculations}

\subsection{Which ring isomer has the lowest ground-state energy?: 
REM versus UHF energies}

\begin{figure}[t]
\centering\includegraphics[width=7.0cm]{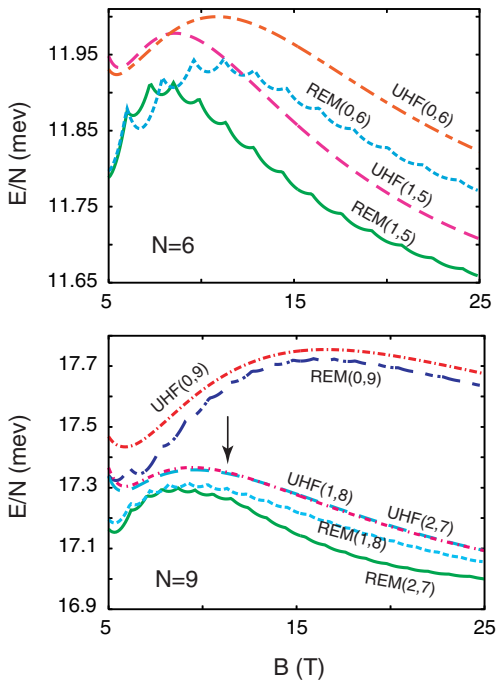}
\caption{(Color online)
Comparison of REM and UHF ground-state energies per particle 
(referenced to $\hbar \Omega$)
associated with different ring isomers for $N=6$ and $N=9$ electrons 
as a function of the magnetic field $B$.
The curves are labeled with the computational method and the
isomer $(n_1,n_2)$. 
To the left of the vertical arrow (at $B=11.5$ T), the UHF(1,8) curve is 
energetically favored.  
To the right of the vertical arrow, the UHF(2,7) curve is energetically
favored.
Parameters used: confinement $\hbar \omega_0=3.60$ meV, 
dielectric constant $\kappa=13.1$, effective mass $m^*=0.067 m_e$. 
}
\end{figure}

For a given number $N$ of electrons, there exist\cite{kon} in general more than
one polygonal-ring isomers, associated with stable and metastable 
equilibrium configurations of $N$ electrons inside an external harmonic
confinemnet $\hbar \omega_0$. Figure 3 displays UHF and REM ground-state 
energies for $N=6$ and $N=9$ electrons associated with the various classical 
polygonal-ring configurations. For $N=6$, one has two isomers, i.e., a (0,6)
configuration and a (1,5) configuration (with one lectron at the center).
For $N=9$ electrons, there exist three different isomers, i.e., (0,9), (1,8), and
(2,7). From the bottom panel in Fig. 3, 
we observe that for $N=9$ electrons, the lowest REM energies 
correspond to the classically stable isomer, i.e., to the (2,7) configuration
with two electrons in the inner ring and seven electrons in the outer ring.   
In particular, we note that the (0,9) isomer (which may be associated with a
single-vortex state) yileds REM energies far above the (2,7) one in the
whole magnetic-field range  5 T $< B <$ 25 T, and in particular for magnetic
fields immediately above those associated with the MDD (the so-called MDD
break-up range); the MDD for $N=9$ electrons has an angular momentum $L_0=36$
and corresponds to the first energy oscillation in the figure. 

We have found that the $(0,N)$ isomer is not associated with REM ground energies
for any magnetic-field range in all cases with $N \geq 7$. The only instance when
the $(0,N)$ configuration is associated with a REM ground-state energy is the
$N=6$ case [see Fig. 3, top frame], where the REM energy of the (0,6) 
configuration provides the ground-state energy in the range 6.1 T $<B<$ 7.7 T,
immediately after the break-up of the MDD. 

For comparison, we have also plotted in Fig.\ 3 the UHF energies as a function
of the magnetic field. Most noticeable is the fact that the REM ground states,
compared to the UHF ones, may result in a different ordering of the isomers.
For example, in the range 5 T $<B<$ 6.1 T, the UHF indicates, by a small  
energetic advantage, the (0,6) as the ground-state 
configuration associated with the MDD, 
while the REM specifies the (1,5) arrangement as the ground-state configuration.
A similar switching of the ground-state isomers is also seen between the (1,8) 
and (2,7) configurations in the case of $N=9$ electrons in the 
magnetic-field range 5 T $<B<$ 11.5 T.
We conclude that transitions between the different electron-molecule isomers 
derived from UHF energies alone\cite{szaf,note32} are not reliable.

\subsection{The case of $N=9$ electrons}

\begin{figure}[t]
\centering\includegraphics[width=6.5cm]{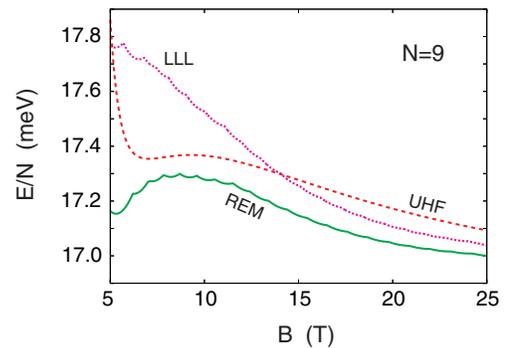}
\caption{(Color online)
Ground-state energies [i.e., for the (2,7) configuration]
for $N=9$ electrons (per particle, referenced to 
$\hbar \Omega$) as a function of the magnetic field $B$. 
Dashed line (red): UHF (SEM). Solid line (blue): REM. Dotted line (black): 
approximate energies 
$\widetilde{E}^{\text{REM}}_{\text{tot,LLL}}(B)$ (see text). 
Parameters used: confinement $\hbar \omega_0=3.60$ meV, 
dielectric constant $\kappa=13.1$, effective mass $m^*=0.067 m_e$. 
}
\end{figure}

In figure 4 we show ground-state energies for the case of $N=9$ 
electrons, which have a nontrivial double-ring configuration $(n_1,n_2)$. 
Here, the most stable configuration for classical point charges\cite{kon}
is $(2,7)$, for which we have carried 
UHF (SEM) and REM (projected) calculations in the magnetic field range 
5 T $< B <$ 25 T. We also display in Fig.\ 4 the energies 
$\widetilde{E}^{\text{REM}}_{\text{tot,LLL}}(B)$ [dotted curve (black)], which, 
as in the $N=4$ and $N=3$ cases discussed in the section III, overestimate the 
ground-state energies, in particular for smaller $B$. \cite{note39}
In keeping with the findings for
smaller sizes \cite{yl}$^{(c)}$ [with $(0,N)$ or $(1,N-1)$ configurations], both 
the UHF and the REM ground-state energies of the $N=9$ case approach
as $B \rightarrow \infty$ the {\it classical\/} equilibrium energy of the (2,7) 
polygonal configuration [i.e., 16.75 meV; 4.088$E_0$ in the units of 
Ref.\ \onlinecite{kon}, $E_0 \equiv (m^* \omega_0^2 e^4/2\kappa^2)^{1/3}$]. 

In analogy with smaller sizes [see, e.g., Figs.\ 1 and 2 for $N=4$ and $N=3$],
the REM ground-state energies in Fig.\ 4 exhibit oscillations as a function
of $B$. These oscillations reflect the incompressibility of the many-body states 
associated with magic angular momenta. The magic angular momenta are specified
by the number of electrons on each ring, and in general they are given by 
$L_m = N(N-1)/2 + \sum_{q=1}^r k_q n_q$, where the $n_q$'s are the number of 
electrons located in the $q$th ring and the $k_q$'s are non-negative
integers; in particular, $L_m=36+2k_1+7k_2$ for the $N=9$ case in Fig.\ 4. 
An analysis of the actual values taken by the set of
indices $\{k_1,k_2 \}$ reveals several additional trends that further
limit the allowed values of ground-state $L_m$'s. In particular, starting
with the values $\{k_1=0,k_2=0 \}$ at $B=5$ T ($L_m^{\text{MDD}}=36$), the
indices $\{k_1,k_2 \}$ reach the values $\{2,24 \}$ at $B=25$ T
($L_m=208$). As seen form Table II, the outer index $k_2$ has a short period,
while the inner index $k_1$ exhibits a longer period and increases much
more slowly than $k_2$. This behavior minimizes the total kinetic energy of the
independently rotating rings (having a variable radius, see section V below).

\begin{table}[t]
\caption{%
Ground-state magic angular momenta and their decomposition $\{ k_1,k_2\}$
for $N=9$ in the nagnetic-field range 5 T $ \leq B \leq $ 25 T. These results
correspond to the REM [see lower curve in Fig.\ 4, with the electrons
arranged in a (2,7) structure]. The parameters
used are as in Fig.\ 4.
}
\begin{ruledtabular}
\begin{tabular}{rrr|rrr}
$L_m$  & $k_1$ & $k_2$ & $L_m$  & $k_1$ & $k_2$    \\ \hline
36  & 0 & 0 & 129 & 1  & 13  \\
43  & 0 & 1 & 136 & 1  & 14  \\ 
50  & 0 & 2 & 143 & 1  & 15  \\
57  & 0 & 3 & 150 & 1  & 16  \\
64  & 0 & 4 & 157 & 1  & 17  \\
71  & 0 & 5 & 164 & 1  & 18  \\
78  & 0 & 6 & 171 & 1  & 19  \\
87  & 1 & 7 & 173 & 2  & 19  \\
94  & 1 & 8 & 180 & 2  & 20  \\
101  & 1 & 9 & 187 & 2  & 21  \\
108  & 1 & 10 & 194 & 2 & 22   \\
115  & 1 & 11 & 201 & 2 & 23  \\
122  & 1 & 12 & 208 & 2 & 24  
\end{tabular}
\end{ruledtabular}
\end{table}

\begin{table}[b]
\caption{%
Ground-state magic angular momenta and their decomposition $\{ k_1,k_2\}$
for $N=9$ electrons associated with the 
$\widetilde{E}_{\text{tot,LLL}}^{\text{REM}}$ curve
[top curve in Fig.\ 4; see section II.B for an explanation of notation;
the electrons are arranged in a (2,7) structure].
}
\begin{ruledtabular}
\begin{tabular}{rrr|rrr}
$L_m$  & $k_1$ & $k_2$ & $L_m$  & $k_1$ & $k_2$    \\ \hline
36  & 0 & 0 & 57 & 0  & 3  \\
45  & 1 & 1 & 64 & 0  & 4  \\ 
52  & 1 & 2 & 71 & 0  & 5  
\end{tabular}
\end{ruledtabular}
\end{table}

We also list in Table III the first few pairs of indices $\{ k_1,k_2\}$
associated with the $\widetilde{E}_{\text{tot,LLL}}^{\text{REM}}$ 
curve (see top dotted curve in Fig.\ 4). It can be seen that the magic angular 
momenta are different from those associated with the REM curve, when the
confinement is taken into consideration using the full projected 
wave function in Eq.\ (\ref{wfprj}). The magic angular momenta of the
$\widetilde{E}_{\text{tot,LLL}}^{\text{REM}}$ curve coincide 
with the $L_m$'s of the EXD within
the LLL, and thus are characterized by having $L_0+N=45$ 
(instead of $L_0+n_2=43$) as the magic 
angular momentum immediately following that of the MDD (i.e., $L_0=36$). 
The $L_0+N$ magic angular momentum is associated with the $(0,N)$ ring 
arrangement, which can be interpreted as a single ``vortex-in-the-center'' 
state. 

Based on EXD calculations restricted to the lowest Landau level 
\cite{yang2,nie,reim} (that is, $\widetilde{E}_{\text{int,LLL}}^{\text{EXD}}$ or 
$\widetilde{E}_{\text{tot,LLL}}^{\text{EXD}}$ in our notation), 
it has been conjectured that for QDs with $N < 15$, the break-up of the MDD with 
increasing $B$ proceeds through the formation of the above mentioned 
single central vortex state. However, our REM calculations show 
(see also the case of $N=11$ electrons in 
section IV.C and the case of $N=17$ electrons in section 
IV.D) that taking into account properly the influence of the confinement does 
not support such a scenario. Instead, the break-up of the MDD resembles an edge 
reconstruction and it proceeds (for all dots with $N > 6$) through the gradual 
detachement of the {\it outer ring\/} associated 
with the classical ground-state polygonal configuration  
(see Table II for the case of $N=9$ electrons). The only case we found where the
break-up of the MDD proceeds via a (0,N) vortex state is the one with 
$N=6$ electrons (see section IV.A),; and naturally the cases with $N \leq 5$. 

\begin{figure}[t]
\centering\includegraphics[width=7.6cm]{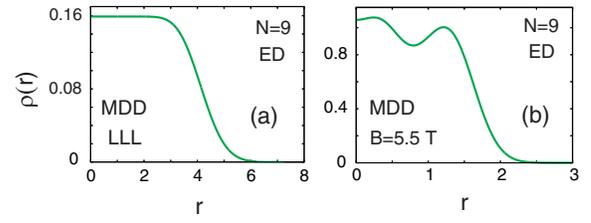}
\caption{(Color online)
REM radial electron densities for the MDD ($L_m=L_0=36$) of $N=9$ electrons 
[in the (2,7) ground-state configuration] at (a) $B \rightarrow \infty$,
i.e., in the lowest Landau level and (b) at $B = 5.5$ T.
Parameters used in (b): 
confinement $\hbar \omega_0=3.60$ meV, dielectric constant $\kappa=13.1$, 
effective mass $m^*=0.067 m_e$. Lengths: (a) in units of the magnetic 
length $l_B$; (b) in units of $R_0=(2e^2/m^* \kappa \omega_0^2)^{1/3}$.
Electron densities: (a) in units of $1/l_B^2$; (b) in units of $1/R_0^2$.
Normalization: $2 \pi \int_0^\infty \rho(r)rdr = N$.
}
\end{figure}

\begin{figure}[b]
\centering\includegraphics[width=8.0cm]{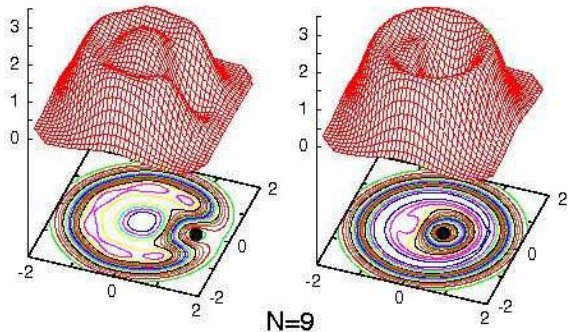}
\caption{(Color online)
Conditional probability distributions obtained from REM wave functions of the 
MDD ($L_0=36$) for $N=9$ electrons at $B=5.5$ T [see Fig. 5(b)].
The electrons are arranged in a (2,7) structure. 
The observation point is denoted by a solid dot. On the left, the
observation point is located on the outer shell, and on the right it is located
on the inner shell.
Parameters used: confinement $\hbar \omega_0=3.60$ meV, 
dielectric constant $\kappa=13.1$, effective mass $m^*=0.067 m_e$. 
Lengths in units of $R_0 =( 2e^2/(\kappa m^* \omega_0^2) )^{1/3}$.
CPDs (vertical axes) in arbitrary units.
}
\end{figure}

As another illustration of the subtle, but important, 
differences that exist between wave functions defined exclusively within the 
LLL and those specified by the REM wave function for finite $B$ in Eq.\ 
(\ref{wfprj}), we display in Fig.\ 5 for $N=9$ electrons
the radial electron densities of the MDD at $B \rightarrow \infty$ 
and at $B=5.5$ T. While the electron density of the MDD 
in the LLL ($B \rightarrow \infty$) is constant in the central region [up
to $r \approx 3 l_B$, see Fig.\ 5(a)], the corresponding density at $B=5.5$ T 
displays the characteristic oscillation corresponding to the (2,7) multi-ring 
structure [see Fig.\ 5(b)]; the latter behavior is due to the
mixing of higher Landau levels. To further illustrate the (2,7) crystalline
character of the MDD when higher Landau levels are considered, we display in 
Fig.\ 6 the corresponding CPDs associated with the REM wave function of the MDD 
at $B=5.5$ T and an external confinement of $\hbar \omega_0= 3.6$ meV.    
Our conclusions concerning the MDD electron densities (and CPDs) are supported 
by EXD calculations for $N=4$ electrons.\cite{note47}
Note that, while the ring structure is well developed in the CPDs
shown in Fig.\ 6, the internal (2,7) structure of the rings (see in particular
the outer ring in the left panel in Fig.\ 6) is rather weak, as expected for the
lowest angular momentum $L_0$ (MDD). However, the ring structure is easily 
discernible in contrast to the CPDs for the MDD 
{\it restricted to the LLL\/} where structureless CPDs (as well as
structureless electron densities) are found.

\subsection{The case of $N=11$ electrons}

\begin{figure}[t]
\centering\includegraphics[width=6.5cm]{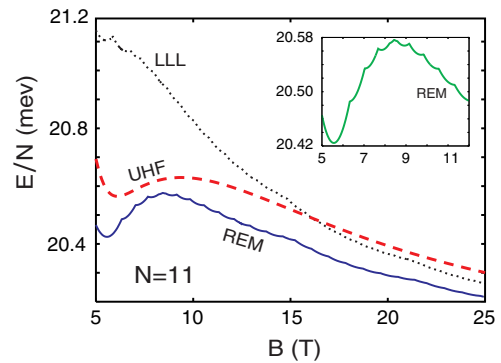}
\caption{(Color online)
Ground-state energies for $N=11$ electrons (per particle, referenced to 
$\hbar \Omega$) as a function of the magnetic field $B$. 
Dashed line (red): UHF (SEM). Solid line (blue): REM. Dotted line (black): 
Approximate energies 
$\widetilde{E}^{\text{REM}}_{\text{tot,LLL}}(B)$ (see text). 
Parameters used: confinement $\hbar \omega_0=3.60$ meV, 
dielectric constant $\kappa=13.1$, effective mass $m^*=0.067 m_e$. 
The inset shows a magnification of the REM curve in the range
5 T $<B<$ 12 T.
}
\end{figure}

\begin{table}[b]
\caption{%
Ground-state magic angular momenta and their decomposition $\{ k_1,k_2\}$
for $N=11$ in the nagnetic-field range 5 T $ \leq B \leq $ 25 T. 
The results correspond to the REM (see lower curve in Fig.\ 7). 
The parameters used are as in Fig.\ 7.
}
\begin{ruledtabular}
\begin{tabular}{rrr|rrr}
$L_m$  & $k_1$ & $k_2$ & $L_m$  & $k_1$ & $k_2$    \\ \hline
55  & 0 & 0 & 165 & 2  & 13  \\
63  & 0 & 1 & 173 & 2  & 14  \\ 
71  & 0 & 2 & 181 & 2  & 15  \\
79  & 0 & 3 & 189 & 2  & 16  \\
90  & 1 & 4 & 197 & 2  & 17  \\
98  & 1 & 5 & 205 & 2  & 18  \\
106  & 1 & 6 & 213 & 2  & 19  \\
114  & 1 & 7 & 224 & 3  & 20  \\
122  & 1 & 8 & 232 & 3  & 21  \\
130  & 1 & 9 & 240 & 3  & 22  \\
138  & 1 & 10 & 248 & 3 & 23   \\
146  & 1 & 11 & 256 & 3 & 24  \\
154  & 1 & 12 &     &   &   
\end{tabular}
\end{ruledtabular}
\end{table}

Figure 7 presents the case for the ground-state energies of 
a QD with $N=11$ electrons, 
which have a nontrivial double-ring configuration $(n_1,n_2)$. The most 
stable\cite{kon} classical configuration is $(3,8)$, for which we have carried 
UHF (SEM) and REM (projected) calculations in the magnetic field range 
5 T $< B <$ 25 T. Figure 7 also displays the LLL ground-state energies 
$\widetilde{E}^{\text{REM}}_{\text{tot,LLL}}(B)$ 
[dotted curve (black)], which, as in previous cases, overestimate
the ground-state energies for smaller $B$. The approximation
$\widetilde{E}^{\text{REM}}_{\text{tot,LLL}}(B)$, however, can be used 
to calculate ground-state energies for higher values of $B$.  
In keeping with the findings for smaller sizes \cite{yl}$^{(c)}$ [with $(0,N)$ 
or $(1,N-1)$ configurations], we found that both 
the UHF and the REM ground-state energies approach, 
as $B \rightarrow \infty$, the {\it classical\/} equilibrium energy of the (3,8) 
polygonal configuration [i.e., 19.94 meV; 4.865$E_0$ in the units of 
Ref.\ \onlinecite{kon}, $E_0 \equiv (m^* \omega_0^2 e^4/2\kappa^2)^{1/3}$]. 

In analogy with smaller sizes [see, e.g., Figs.\ 1, 2, and 4 for $N=4$, 3,
and 9, respectively], the REM ground-state energies in Fig.\ 7 
exhibit oscillations as a function of $B$ (see in particular the inset). 
As discussed in section IV.B, these oscillations are associated with magic 
angular momenta, specified by the number of electrons on each ring. 
For $N=11$ they are given by Eq.\ (\ref{lmeq}), i.e., $L_m=55+3k_1+8k_2$, with 
the $k_q$'s being nonnegative integers. As was the case with $N=9$ electrons, 
an analysis of the actual values taken by the set of
indices $\{k_1,k_2 \}$ reveals several additional trends that further
limit the allowed values of ground-state $L_m$'s. In particular, starting
with the values $\{0,0 \}$ at $B=5$ T ($L_0=55$), the
indices $\{k_1,k_2 \}$ reach the values $\{3,24 \}$ at $B=25$ T
($L_m=256$). As seen from Table IV, the outer index $k_2$ changes faster
than the inner index $k_1$. This behavior minimizes the total kinetic energy of 
the independently rotating rings; indeed, the kinetic energy of the inner ring
(as a function of $k_1$) rises faster than that of the outer ring (as a function
of $k_2$) due to smaller moment of inertia (smaller radius) of the inner ring
[see Eq.\ (\ref{eclkin})]. 
  
\begin{figure}[t]
\centering\includegraphics[width=7.0cm]{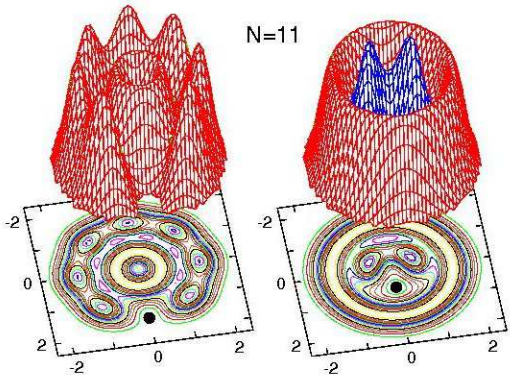}
\caption{(Color online)
REM conditional probability distributions for $N=11$ electrons at $B=10$ T 
($L=106$). The electrons are arranged in a (3,8) structure. The observation point
(solid dot) is placed on (left) the outer ring at $r_0=1.480 R_0$, and 
(right) on the inner ring at $r_0=0.557 R_0$. 
Parameters used: confinement $\hbar \omega_0=3.60$ meV, 
dielectric constant $\kappa=13.1$, effective mass $m^*=0.067 m_e$.
Lengths in units of $R_0=(2e^2/m^* \kappa \omega_0^2)^{1/3}$. 
CPDs (vertical axes) in arbitrary units.
}
\end{figure}

In addition to the overestimation of the ground-state energy values, 
particularly for smaller magnetic fields (see Fig.\ 7 and our above discussion),
the shortcomings of the LLL approximation pertaining to the ground-state ring 
configurations [see section II.B, Eq.\ (\ref{limexd})], as discussed by us
above for $N=9$, persist also for $N=11$.
In particular, we find that according to the LLL approximation the 
ground-state angular-momentum immediately after the MDD ($L_0=55$) is $L_m=66$,
i.e., the one associated with the $(0,N)$ vortex-in-the-center configuration. 
This result, erroneously stated in Ref.\ \onlinecite{reim} as the ground state,
disagrees with the correct result that includes the effect of the
confinement -- listed in Table IV, where the ground-state angular momentum 
immediately following the MDD is $L_m=63$. This angular momentum corresponds
to the classicaly most stable (3,8) ring configuration -- that is a 
configuration with no vortex at all.

Figure 8 displays the REM conditional probability distributions
for the ground state of $N=11$ electrons at $B=10$ T ($L_m=106$). The (3,8) ring
configuration is clearly visible. We note that when the observation point is 
placed on the outer ring (left panel), the CPD
reveals the crystalline structure of this ring only; the inner ring appears to
have a uniform density. To reveal the crystalline structure of the inner ring,
the observation point must be placed on this ring; then the outer ring
appears to be uniform in density. This behavior suggests that the two rings
rotate independently of each other, a property that we will explore in section
V to derive an approximate expression for the yrast rotational spectra associated
with an arbitrary number of electrons.

\subsection{The case of $N=17$ electrons}

\begin{figure}[b]
\centering\includegraphics[width=6.5cm]{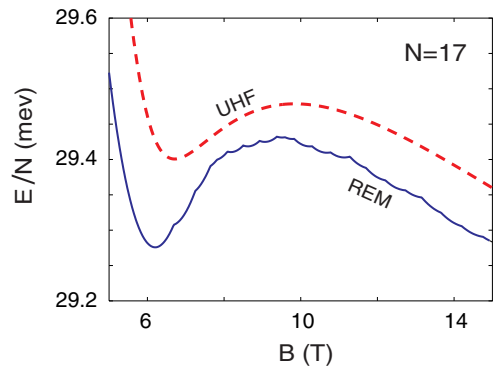}
\caption{(Color online)
Ground-state energies (per particle, referenced to $\hbar \Omega$) for $N=17$ 
electrons as a function of the magnetic field $B$. The electrons are arranged
in a (1,6,10) structure. Dashed line (red): UHF. Solid line (blue): REM. 
Parameters used: confinement $\hbar \omega_0=3.6$ meV, 
dielectric constant $\kappa=13.1$, effective mass $m^*=0.067 m_e$. 
}
\end{figure}

Figure 9 presents (for 5 T $\leq B \leq$ 15 T) REM and UHF ground-state energies
for $N=17$ electrons, which have a (1,6,10) three-ring configuration as the most
stable classical arrangement. \cite{kon}

In analogy with smaller sizes [see, e.g., previous figures for $N \leq 12$]
the REM ground-state energies in Fig.\ 9 exhibit oscillations as a function
of $B$, and each oscillation is associated with a given particular (magic) 
value of the angular momentum. Earlier in this section we discussed the
physical origins of the magic angular momenta. As before, 
the magic angular momenta are specified by the number of electrons on each ring
[(\ref{lmeq})], with $L_0=136$ and $L_m=136+6k_2+10k_3$ for $N=17$; 
$k_q$'s being non-negative integers
(the central electron does not contribute to the total angular momentum). 
Analysis of the particular values taken by the set of indices $\{k_2,k_3 \}$ 
reveals similar trends to those found for the cases with $N=9$ and $N=11$ 
electrons. In particular, starting
with the values $\{0,0 \}$ at $B=5$ T ($L_0=136$), the
indices $\{k_2,k_3 \}$ reach the values $\{k_2=5,k_3=18 \}$ at $B=15$ T
($L_m=346$). As seen from Table V, the outer index $k_3$ changes faster,
than the inner index $k_2$. This behavior minimizes the total kinetic energy of
the independently rotating rings, as was already discussed for $N=9$ and $N=11$ 
electrons. 

We have also calculated the ground-state energies for $N=17$ electrons
in the LLL approximation, i.e., we calculated the quantity
$\widetilde{E}^{\text{REM}}_{\text{tot,LLL}} (B)$ (not shown in Fig.\ 9). We find
once more that $\widetilde{E}^{\text{REM}}_{\text{tot,LLL}} (B)$ 
overestimates the
ground-state energies in the magnetic-field range covering the MDD and the 
range immediately above the MDD. For $N=17$, however, the shortcoming of
the LLL approximation is not reflected in the determination of the
ground-state ring configurations. We find that for $N=17$ the LLL approximation
yields a (1,6,10) ring configuration (with $L_m=146$) for the ground state
immediately following the MDD, in agreement with the REM 
configurations listed in Table V.
\begin{table}[t]
\caption{%
Ground-state magic angular momenta and their decomposition $\{ k_2,k_3\}$
for $N=17$ electrons in the nagnetic-field range 5 T $ \leq B \leq $ 15 T. 
The rersults correspond to the REM (see lower curve in Fig.\ 9).
The parameters used are as in Fig.\ 9.
}
\begin{ruledtabular}
\begin{tabular}{rrr|rrr}
$L_m$  & $k_2$ & $k_3$ & $L_m$  & $k_2$ & $k_3$    \\ \hline
136  & 0 & 0 & 238 & 2  & 9    \\
146  & 0 & 1 & 248 & 2  & 10   \\
156  & 0 & 2 & 264 & 3  & 11   \\
166  & 0 & 3 & 274 & 3  & 12   \\
172  & 1 & 3 & 284 & 3  & 13   \\
182  & 1 & 4 & 294 & 3  & 14   \\
192  & 1 & 5 & 310 & 4  & 15   \\
202  & 1 & 6 & 320 & 4  & 16   \\
212  & 1 & 7 & 330 & 4  & 17   \\
218  & 2 & 7 & 340 & 4  & 18   \\
228  & 2 & 8 & 346 & 5  & 18
\end{tabular}
\end{ruledtabular}
\end{table}

\begin{figure}[b]
\centering\includegraphics[width=8.cm]{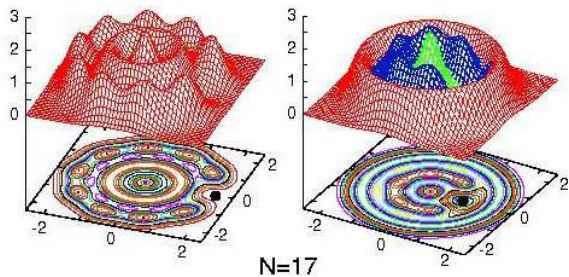}
\caption{(Color online)
Grond-state conditional probability distributions, CPDs, obtained from REM wave 
functions for the ground state of $N=17$ electrons at $B=10$ T ($L=228$).
The electrons are arranged in a (1,6,10) structure.
The observation point (solid dot) is placed on the outer ring at
$r_0=1.858R_0$ (left frame), and on the inner ring at $r_0=0.969R_0$
(right frame). The rest of the parameters are the same as in Fig.\ 9.
Lengths in units of $R_0 =( 2e^2/(\kappa m^* \omega_0^2) )^{1/3}$.
CPDs (vertical axes) in arbitrary units.
}
\end{figure}

\section{REM YRAST BAND EXCITATION SPECTRA AND DERIVATION OF ANALYTIC
APPROXIMATE FORMULA} 

In Fig.\ 10, we display the CPD for the REM wave function of $N=17$ electrons.
This case has a nontrivial three-ring structure (1,6,10), \cite{kon}
which is sufficiently complex to allow generalizations for larger numbers of
particles. The remarkable combined character (partly crystalline
and partly fluid leading to a non-classical rotational inertia, see section VI) 
of the REM is illustrated in the CPDs of Fig.\ 10. Indeed, as the two CPDs
[reflecting the choice of taking the observation point [${\bf r}_0$ in
Eq.\ (\ref{cpds})] on the outer (left frame) or the inner ring (right frame)]
reveal, the polygonal electron rings rotate {\it independently\/} of each other.
Thus, e.g., to an observer located on the inner ring, the outer ring will appear
as having a uniform density, and vice versa. The wave functions obtained from
exact diagonalization exhibit also the property of independently rotating rings
[see e.g., the $N=12$ and $L=132$ ($\nu=1/2$) case in Fig.\ 11], which
is a testimony to the ability of the REM wave function to capture the essential
physics \cite{jain} of a finite number of electrons in high $B$. In particular,
the conditional probability distribution obtained for exact diagonalization
wave functions in Fig.\ 11 exhibits the characteristics expected from the CPD 
evaluated using REM wave functions for the (3,9) configuration and with an 
angular-momentum decomposition into shell contributions 
[see Eqs.\ (\ref{wfprj}) and (\ref{lmpar})]
$L_1=3+3k_1$ and $L_2=63+9k_2$ ($L_1+L_2=L_m$; for $L_m=132$ the 
angular-momentum decomposition is $L_1=6$ and $L_2=126$).

\begin{figure}[t]
\centering\includegraphics[width=8.0cm]{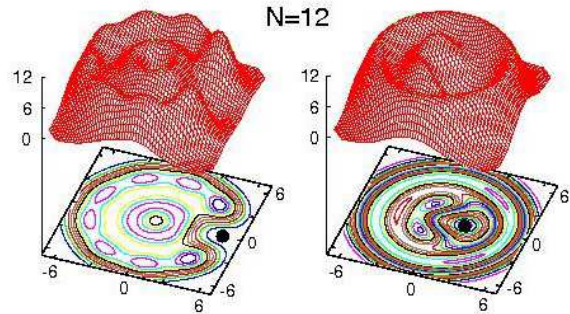}
\caption{(Color online)
CPDs for $N=12$ electrons and with angular momentum $L=132$ ($\nu=1/2$) 
calculated with EXD in the lowest Landau level.
The electrons are arranged in a (3,9) structure.
The observation point (solid dot) is placed on the outer ring at
$r_0=5.22 l_B$ (left frame), and on the inner ring at $r_0=1.87 l_B$
(right frame).  Lengths in units of $l_B$. 
CPDs (vertical axes) in arbitrary units.
}
\end{figure}

In addition to the conditional probabilities, the solid/fluid character of the
REM is revealed in its excited rotational spectrum for a given $B$. From our
microscopic calculations based on the wave function in Eq.\ (\ref{wfprj}), we
have derived (see below) an approximate (denoted as ``app''), but
{\it analytic\/} and {\it parameter-free\/}, expression [see Eq.\
(\ref{app}) below] which reflects directly the nonrigid (nonclassical)
character of the REM for arbitrary size. This expression allows calculation of
the energies of REMs for arbitrary $N$, given the corresponding equilibrium
configuration of confined classical point charges.

We focus on the description of the yrast band at a given $B$.
Motivated by the aforementioned nonrigid character of the rotating electron
molecule, we consider the following kinetic-energy term corresponding to a
$(n_1,...,n_q,...,n_r)$ configuration (with $\sum_{q=1}^r n_q =N$):
\begin{equation}
E_{\text{app}}^{\text{kin}} (N)=
\sum_{q=1}^r \hbar^2 L_q^2/(2 {\cal J}_q (a_q)) - \hbar \omega_c L/2,
\label{eclkin}
\end{equation}
where $L_q$ is the partial angular momentum associated with the $q$th ring 
about the center of the dot and the total angular momentum is 
$L=\sum_{q=1}^r L_q$. ${\cal J}_q (a_q)) \equiv n_q m^* a_q^2$ is the 
rotational moment of inertia of each {\it individual\/} ring, 
i.e., the moment of inertia
of $n_q$ classical point charges on the $q$th polygonal ring of radius $a_q$.
To obtain the total energy, $E_L^{\text{REM}}$, we include also the term
$E_{\text{app}}^{\text{hc}} (N) =\sum_{q=1}^r  {\cal J}_q (a_q) \Omega^2/2$
due to the effective harmonic confinement $\Omega$ [see discussion of Eq.\
(\ref{uhfo})], as well as the interaction energy $E_{\text{app}}^C$,
\begin{equation}
E_{\text{app}}^C (N) = \sum_{q=1}^r \frac{n_q S_q}{4} \frac{e^2}{\kappa a_q} +
\sum_{q=1}^{r-1} \sum_{s > q}^r V_C(a_q,a_s).
\label{vc}
\end{equation}
The first term is the intra-ring Coulomb-repulsion energy of $n_q$ point-like 
electrons on a given ring, with a structure factor
\begin{equation}
S_q = \sum_{j=2}^{n_q}(\sin[(j-1)\pi/n_q])^{-1}.
\label{sq}
\end{equation}
The second term is the inter-ring Coulomb-repulsion energy between rings of
uniform charge distribution corresponding to the specified numbers 
of electrons on the polygonal rings. The expression fo $V_C$ is
\begin{eqnarray}
V_C(a_q,a_s)&=& n_q {n_s}\; {_2F_1} [3/4,1/4;1;4 a_q^2 a_s^2(a_q^2+a_s^2)^{-2}]
\nonumber \\
&& \times e^2 (a_q^2+a_s^2)^{-1/2}/\kappa, 
\label{vcc}
\end{eqnarray}
where ${_2F_1}$ is the hypergeometric function.

For large $L$ (and/or $B$), the radii of the rings of the rotating molecule
can be found by neglecting the interaction term in the total approximate energy,
thus minimizing only
$E_{\text{app}}^{\text{kin}} (N) + E_{\text{app}}^{\text{hc}} (N)$.
One finds 
\begin{equation}
a_q = \lambda \sqrt{L_q/n_q};
\label{rad}
\end{equation}
i.e., the ring radii depend on the partial angular momentum $L_q$,
reflecting the {\it lack of radial rigidity\/}. Substitution into the above
expressions for $E_{\text{app}}^{\text{kin}}$, $E_{\text{app}}^{\text{hc}}$, and
$E_{\text{app}}^C$ yields for the total approximate energy the final
expression:
\begin{eqnarray}
&E_{\text{app},L}^{\text{REM}}&(N) =  \hbar(\Omega-\omega_c/2) L + \nonumber \\
&& \sum_{q=1}^r \frac{C_{V,q}}{L_q^{1/2}} +
\sum_{q=1}^{r-1} \sum_{s > q}^r
V_C(\lambda \sqrt{\frac{L_q}{n_q}}, \lambda \sqrt{\frac{L_s}{n_s}}),
\label{app}
\end{eqnarray}
where the constants
\begin{equation}
C_{V,q}=0.25 n_q^{3/2} S_q e^2/(\kappa \lambda).
\label{ccvq}
\end{equation}
For simpler $(0,N)$ and $(1,N-1)$ ring configurations,
Eq.\ (\ref{app}) reduces to the expressions reported earlier.$^{7(c),}$
\cite{mak}

\begin{figure}[t]
\centering\includegraphics[width=7.5cm]{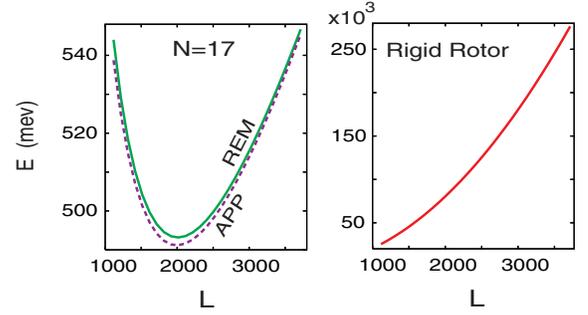}
\caption{(Color online)
Left: Yrast spectrum for $N=17$ electrons at a high magnetic field $B=100$ T.
Approximate analytic expression [Eq.\ (\ref{app}), dashed line (violet)]
compared with microscopic REM calculations [Eq.\ (\ref{eproj}), solid
line (green)].
Right: The corresponding classical (rigid rotor) energy $E^{\text{rig}}_L$
for $N=17$ electrons (see text). The microscopic REM energies are referenced
relative to the zero-point energy, $17 \hbar \Omega$. Energies
were calculated for magic angular momenta
$L=L_1+L_2+L_3$ with $L_1=0$, $L_2=21+6k_2$ and $k_2=30$, and $L_3=115+10k_3$.
The parameters are the same as in Fig.\ 9.
Note the much larger energy scale for the classical case (right frame),
leading to a non-rigidity index for the REM of $\alpha \sim
0.99$ (see text).
}
\end{figure}

\section{A non-rigid crystalline phase: Non-classical rotational inertia of
electrons in quantum dots}

In Fig.\ 12 (left frame), and for a sufficiently high magnetic field
(e.g., $B=100$ T such that the Hilbert space of the system reduces to the
lowest Landau level), we compare the approximate analytic energies
$E_{\text{app},L}^{\text{REM}}$ with the microscopic energies
$E_L^{\text{REM}}$ calculated from Eq.\ (\ref{eproj}) using the same
parameters as in Fig. 9 for $N=17$ electrons. The two calculations agree well,
with a typical difference of less than 0.5\% between them. 
More important is the marked difference between these results and the total 
energies of the {\it classical\/} (rigid rotor) molecule ($E^{\text{rig}}_L$), 
plotted in the right frame of Fig.\ 12; the latter
are given by 
\begin{eqnarray}
E^{\text{rig}}_L &=& \hbar^2 L^2 / (2 {\cal J}_{\text{rig}})
+ 0.5 \sum_{i=1}^N m^* \omega_0^2 |Z_i|^2 + \nonumber \\
&&  \sum_{i=1}^N \sum_{j > i}^N e^2/ (\kappa |Z_i-Z_j|),
\label{erig}
\end{eqnarray} 
with the rigid moment of inertia being \cite{note3}
\begin{equation}
{\cal J}_{\text{rig}}=\sum_{i=1}^N m^* |Z_i|^2. 
\label{mirig}
\end{equation}

The disagreement between the REM and the classical energies is twofold: (i)
The $L$ dependence is different, and (ii) The REM energies are three orders of
magnitude smaller than the classical ones. That is, the energy cost for the
rotation of the REM is drastically smaller than for the classical rotation, thus
exhibiting non-classical rotational behavior. 
In analogy with Ref.\ \onlinecite{legg}, we define a ``non-rigidity'' index
\begin{equation}
\alpha = (E^{\text{rig}}_L - E_L^{\text{REM}})/E^{\text{rig}}_L. 
\label{sindex}
\end{equation}
For the case displayed in Fig.\ 12, we find that this index varies
(for $1116 \leq L \leq 3716$) from $\alpha = 0.978$ to $\alpha = 0.998$,
indicating that the rotating electron molecule, while possessing crystalline
correlations is (rotationally) of a high non-rigid nature. We remark that
our definition of $\alpha$ in Eq.\ (\ref{sindex}) was motivated by a similar
form of an index of supefluid fraction introduced in Ref.\ \onlinecite{legg};
we do not mean to imply the presence of a superfluid component for
electrons in quantum dots.
 
In the context of the appearance of supersolid behavior of $^4$He under
appropriate conditions, formation of a supersolid fraction is often discussed in
conjunction with the presence of (i) real defects and (ii) real vacancies. 
\cite{anlf,ches} Our REM wave function [Eq.\ \ref{wfprj}] belongs to a third 
possibility, namely to {\it virtual\/} defects and vacancies, with the number of
particles equal to the number of lattice sites (in the context of $^4$He, the 
possibility of a supersolid with equal
number of particles and lattice sites is mentioned in Ref.\ \onlinecite{legg2}).
Indeed, the azimuthal shift of the electrons by
$(\gamma_1, \gamma_2,...,\gamma_r)$ [see Eq.\ (\ref{wfprj})] may be viewed
as generating {\it virtual\/} defects and vacancies with respect to the original
electron positions at $(\gamma_1=0, \gamma_2=0,...,\gamma_r=0)$ on the
polygonal rings.

A recent publication \cite{jeon} has explored the quantal nature of the
2D electron molecules in the lowest Landau level ($B \rightarrow \infty$)
using a modification of the second-quantized LLL form of the REM wave
functions. \cite{yl} In particular, the modification
consisted of a multiplication of the {\it parameter free\/} REM wave function by
variationally adjustable Jastrow-factor vortices. Without consideration of the
rotational properties of the modified wave function, the inherently quantal
nature of the molecule was attributed exclusively to the Jastrow factor.
However, as shown above, the original REM wave function [Eq.\ (\ref{wfprj})]
already exhibits a characteristic non-classical rotational inertia (NCRI).
Consequently, the additional {\it variational\/} freedom introduced by the
Jastrow prefactor may well lead energetically to a slight numerical improvement,
but it does not underlie the {\it essential quantal\/} physics of the system.
Indeed, as discussed previously and illustrated in detail above,
the important step is the projection of the static electron molecule
onto a state with good total angular momentum [see Eqs. (\ref{uhfo}) and
(\ref{wfprj})].

\section{Summary}

The focus of this study pertains to the development of methods that permit
investigations of the energetic, structural, and excitation properties of 
quantum dots in strong magnetic fields with an (essentially) arbitrary number of 
electrons. Towards this aim, we utilized several computational methods, and have
assessed their adequacy. The methods that we have used are: (1) Exact
diagonalization which is limited to a rather small number of particles; (2) 
The ``two-step'' successive-hierarchical-approximations method 
(see section II.A), in which a UHF step leading to broken-symmetry solutions
(static electron molecule) is followed by restoration (via projection 
techniques) of circularly symmetric states with good angular momenta
(rotating electron molecule; REM); (3) An approximation method based on
diagonalization of the electron-electron interaction term restricted to the
lowest Landau level (LLL). In this method, the total energy includes, in addition
to the LLL diagonalization term, a contribution from the harmonic confinement
that is linear in the total angular momentum; (4) An analytic expression 
[see section V, Eq.\ (\ref{app})] whose derivation is based on the REM.

We performed comparative calculations for quantum dots with an increasing number
of parabolically confined electrons ($N=3$, 4, 6, 9, 11, and 17). The 
ground-state arrangements of the electrons become structurally
more complex as the number of electrons in the dot increases. 
Using the notation $(n_1, n_2, n_3, ...)$ for the number of electrons located
on concentric polygonal rings (see section II.A), the ground-state arrangements 
are: (0,3) for $N=3$, (0,4) for $N=4$, (1,5) for $N=6$, (2,7) for $N=9$,
(3,8) for $N=11$, and (1,6,10) for $N=17$.

Analysis of the results of our calculations revealed that, for all sizes studied 
by us, the two-step REM method provides a highly accurate description of 
electrons parabolically confined in quantum dots for a whole range of applied
magnetic fields, starting from the neighborhood of the so-called maximum
density droplet and extending to the $B \rightarrow \infty$ limit. In contrast,
the LLL-diagonalization approximation was found to be rather inaccurate for 
weaker magnetic fields, where it grossly overestimates the total energies of the 
electrons; the accuracy of this latter method improves at higher field strengths.

The ground-state energy of the electrons in a QD oscillates as a function
of the applied magnetic field, and the allowed values of the angular momenta
are limited to a set of magic angular momentum values, $L_m$, which are a 
natural consequence of the geometrical arrangement of the electrons in the
rotating electron molecule. Accordingly, the electrons are localized on
concentric polygonal rings which rotate independently of each other
(as observed from the conditional probability distributions, see section IV).
Underlying the aforementioned oscillatory behavior is the incompressibility
of the many-body states associated with the magic angular momenta. The general
expression for $L_m$ is given in Eq.\ (\ref{lmeq}), for a given number $N$
and occupancy of the polygonal rings $\{ n_q \}$. For the ground-state 
$L_m$'s, the values of the non-negative integers $k_q$ in Eq.\ (\ref{lmeq})
are taken such as to minimize the total kinetic energy of the electrons.
Since the moment of inertia of an outer ring is larger than that of an inner 
ring of smaller radius, the rotational energy of the outer ring will 
increase more slowly with increasing angular momentum. Therefore, the
$k_q$ index in Eq.\ (\ref{lmeq}) of an outer ring will very up to relatively
large values while the values corresponding to inner rings remain small
(see section IV). As a consequence, we find (see section IV.B to IV.D)
through REM calculations with proper treatment of the confining potential
that for $N > 6$, with increasing strength of the magnetic field, the
maximum density droplet converts into states with no central vortex, in
contrast to earlier conclusions\cite{nie,yang2,reim} drawn on the basis
of approximate calculations restricted to the lowest Landau level.
Instead we find that the break-up of the MDD with increasing $B$ proceeds 
through the gradual detachment of the outer ring associated with the 
corresponding classical polygonal configuration.

In addition to the ground-state geometric arrangements, we have studied for 
certain sizes higher-energy structural isomers (see, e.g., the cases of $N=6$ 
and $N=9$ confined electrons in Fig.\ 3). We find that for all cases with 
$N \geq 7$ multi-ring confined-electron structures $(n_1, n_2, ..., n_r)$, with 
$n_1, n_2, ...,n_r \neq 0$ and $r \geq 2$, are energetically favored. 
For $N=6$, a (1,5) structure is favored except for a small $B$-range 
(e.g., 6.1 T $< B< $ 7.7 T for the parameters in Fig.\ 3), where the (0,6) 
single-ring structure is favored. For $N \leq 5$ the $(0,N)$ single-ring 
structure is favored for all $B$ values. 

In the REM calculations, 
we have utilized an analytic many-body wave function [Eq.\ (\ref{wfprj})]
which allowed us to carry out computations for a sufficiently large
number of electrons ($N=17$ electrons having a nontrivial three-ring
polygonal structure), leading to the derivation and validation of an analytic
expression Eq. (\ref{app}) for the total energy of rotating electron 
crystallites of arbitrary $N$. 

The non-rigidity implied by the aforementioned independent rotations of the 
individual concentric polygonal rings motivated us to quantify (see section
VI) the degree of non-rigidity of the rotating electron molecules at high
$B$, in analogy with the concept of non-classical rotational
inertia used in the analysis\cite{legg,legg2} of supersolid $^4$He.
These findings for finite dots suggest a strong quantal nature for 
the extended Wigner crystal in the lowest Landau level, designating it 
as a useful paradigm for exotic quantum solids.\\
~~~~\\

\acknowledgments

This research is supported by the U.S. D.O.E. (Grant No. FG05-86ER45234)
and the NSF (Grant No. DMR-0205328).

\appendix

\section{Proof that $u(z,Z)$ [Eq.\ (\ref{uhfo})] lies in the LLL when 
$\lambda= l_B \sqrt{2}$}

Using the identity $-i(xY-yX) = (zZ^*-z^*Z)/2$, one finds
\begin{eqnarray}
&& u(z,Z; \lambda = l_B \sqrt{2}) = \nonumber \\ 
&& \frac { e^\frac{-zz^*-ZZ^*+2zZ^*}{4l_B^2} } {\sqrt{2\pi} l_B} =
\frac{ e^\frac{-zz^*-ZZ^*}{4l_B^2} } {\sqrt{2\pi} l_B} 
\sum_{l=0}^\infty \frac{1}{l!} \left (\frac{zZ^*}{2l_B^2} \right)^l = \nonumber\\
&& \sum_{l=0}^\infty C_l(Z^*) \psi_l(z),
\end{eqnarray}
where $z=x+iy$, $Z=X+iY$, and
\begin{equation} 
C_l(Z^*) = \frac{1}{\sqrt{l!}} 
\left(\frac{Z^*}{l_B \sqrt{2}}\right)^l e^\frac{-ZZ^*}{4l_B^2}, 
\end{equation}
with
\begin{equation}
\psi_l(z)= \frac{1}{\sqrt{2\pi l!}l_B} \left(\frac{z}{l_B \sqrt{2}}\right)^l
e^\frac{-zz^*}{4l_B^2}
\end{equation}
being the Darwin-Fock single-particle wave functions with zero nodes 
forming the LLL.

\section{Coulomb matrix elements between displaced Gaussians [Eq.\ (\ref{uhfo})]}

We give here the analytic expression for the Coulomb matrix elements,
\begin{equation}
V_{ijkl} = \int \int d{\bf r}_1 d{\bf r}_2
u_i^*({\bf r}_1) u_j^*({\bf r}_2) \frac{e^2}{\kappa |{\bf r}_1 - {\bf r}_2|}
u_k({\bf r}_1) u_l({\bf r}_2),
\label{defvijkl}
\end{equation}
between displaced Gaussians [see Eq.\ (\ref{uhfo})] centered
at four arbitrary points $Z_i$, $Z_j$, $Z_k$, and $Z_l$.

One has
\begin{equation}
V_{ijkl} = \frac{e^2}{\kappa \lambda} \sqrt{\frac{\pi}{2}}
e^\vartheta e^{-\varpi} I_0(\varpi),
\label{fvijkl}
\end{equation}
with
\begin{equation}
\vartheta = -\frac{Z_i Z_i^* + Z_j Z_j^* + Z_k Z_k^* + Z_l Z_l^*}{2 \lambda^2}
+\zeta \eta + \sigma \tau,
\label{theta}
\end{equation}
and 
\begin{equation}
\varpi = (\zeta - \sigma)(\eta-\tau)/4,
\label{pi}
\end{equation}
where
\begin{equation}
\zeta = \frac{Z_k + Z_i}{2\lambda} + \beta \frac{Z_k-Z_i}{2\lambda} 
\end{equation}
\begin{equation}
\eta = \frac{Z_i^* + Z_k^*}{2\lambda} + \beta \frac{Z_i^*-Z_k^*}{2\lambda} 
\end{equation}
\begin{equation}
\sigma = \frac{Z_l + Z_j}{2\lambda} + \beta \frac{Z_l-Z_j}{2\lambda} 
\end{equation}
\begin{equation}
\tau = \frac{Z_j^* + Z_l^*}{2\lambda} + \beta \frac{Z_j^*-Z_l^*}{2\lambda}.
\end{equation}

The magnetic-field dependence is expressed through the parameter 
\begin{equation}
\beta = \frac{\lambda^2}{2 l_B^2}.
\end{equation}
The length parameters $\lambda$ and $l_B$ (magnetic length) are defined
in the text following Eq.\ (\ref{uhfo}). Note that $\beta =0$ for $B=0$
and $\beta = 1$ for $B \rightarrow \infty$. The latter offers an
alternative way for calculating REM energies and wave functions in
the lowest Landau level without using the {\it analytic\/} REM wave functions
presented in Ref.\ \onlinecite{yl}.


\begin{thebibliography}{99}
\bibitem{note1}
The REM energies are slightly lower than the EXD ones in several subranges. 
According to the Rayleigh-Ritz variational theorem, this 
indicates that the hyperspherical-harmonics calculation of Ref.\ 
\onlinecite{ruan2} did not converge fully in these subranges.
\bibitem{note32}
In addition, Ref.\ \onlinecite{szaf} performs the UHF variation within the LLL. 
As a result, the electron-molecule configurations found therein beyond the MDD 
resemble those appearing in EXD calculations in the LLL. For $N \leq 14$, 
however, proper consideration of higher Landau levels yields different 
polygonal-ring configurations, as we discuss in detail in sections IV.B and IV.C.
\bibitem{note3}
For the classical equilibrium structure of $N=17$ point charges, the inner
radius is $0.9696R_0$, while the outer one is $1.8418R_0$.\cite{kon}
\end{thebibliography}
\end{document}